\documentclass[aps,prl,twocolumn,superscriptaddress,showpacs]{revtex4}

\usepackage{graphicx}
\usepackage{color}

\begin{document}

\title{Absence of Conventional Spin-Glass Transition in the Ising Dipolar System LiHo$_x$Y$_{1-x}$F$_4$}

\author{P. E. J\"onsson}
\affiliation{Department of Physics, Uppsala University, Box 530, SE-751 21 Uppsala, Sweden}

\author{R. Mathieu}
\affiliation{Department of Microelectronics and Applied Physics, KMF, Royal Institute of Technology (KTH), Electrum 229, SE-164 40 Kista, Sweden}

\author{W. Wernsdorfer}
\affiliation{Institut N\'eel, CNRS/UJF, 25 avenue des Martyrs, BP166, 38042 Grenoble Cedex 9, France}

\author{A. M. Tkachuk}
\affiliation{All-Russia S.I. Vavilov State Optical Institute, St. Petersburg 199034, Russia}

\author{B. Barbara}
\affiliation{Institut N\'eel, CNRS/UJF, 25 avenue des Martyrs, BP166, 38042 Grenoble Cedex 9, France}

\date{\today}


\begin{abstract}
The magnetic properties of single crystals of LiHo$_x$Y$_{1-x}$F$_4$ with $x$=16.5\% and $x$=4.5\% were recorded down to 35 mK using a micro-SQUID magnetometer. While this system is considered as the archetypal quantum spin glass, the detailed analysis of our magnetization data indicates the absence of a phase transition, not only in a transverse applied magnetic field, but also without field. A zero-Kelvin phase transition is also unlikely, as the magnetization seems to follow a non-critical exponential dependence on the temperature. Our analysis thus unmasks the true, short-ranged nature of the magnetic properties of the LiHo$_x$Y$_{1-x}$F$_4$ system, validating recent theoretical investigations suggesting the lack of phase transition in this system.
\end{abstract}

\maketitle
The system LiHo$_x$Y$_{1-x}$F$_4$ ($x$ $\leq$ 0.25) in a transverse field has been considered as the textbook example of the realization of the quantum Ising spin-glass model\cite{aeppli-prb,textbook}. Indeed, a huge uniaxial crystal-field anisotropy gives a strong Ising character to the system, and dipolar\cite{hansen75} interactions between Ho$^{3+}$ ions insure the presence of magnetic couplings of different signs. Furthermore, this glassy system can be obtained in the form of high quality single crystals where Ho$^{3+}$ ions randomly substitute Y$^{3+}$ ones without any modification of the structure (body centered tetragonal lattice with scheelite structure, space group $I4_1/a$\cite{thoma}), so that its intrinsic static and dynamical magnetic properties can be investigated. The quantum and classical spin-glass transitions of these Ho:LiYF$_4$ alloys (with or without transverse field) have been extensively studied in the past, particularly for $x$ = 0.167 and 0.045\cite{aeppli-prb,aeppli-reich,aeppli-wu,aeppli-science,aeppli-nature}. However, the divergence of the non-linear susceptibility $\chi_{nl}$ \cite{phrase} confirming the occurrence of the spin-glass phase transition and allowing the determination of the spin-glass critical exponent $\gamma$ \cite{suzuki,barbara}, has never been analyzed in detail. This sceptical view is also supported by the fact that the ferromagnetic correlation length of the much simpler LiHoF$_4$ in a transverse field is dramatically quenched by hyperfine interaction of the Ho$^{3+}$ ions\cite{ronnow-science}. The generic influence of these interactions on the quantum dynamics of Ho:LiYF$_4$ has been previously demonstrated\cite{giraud}.

Recent theoretical developments predict that due to the presence of off-diagonal hyperfine\cite{schechter-hyperf} and dipolar\cite{schechter-hyperf,schechter-dip,snider,gingras} terms, the Hamiltonian becomes equivalent to that of a ferromagnet with random fields. This is particularly true at low, but non-zero, transverse field where quantum fluctuations are small while the effective random field can be appreciable\cite{schechter-new}.

In this letter, we show that the linear and non-linear susceptibility of high-quality single crystals of LiHo$_x$Y$_{1-x}$F$_4$ with $x$ = 0.165 and 0.045, accurately determined from micro-SQUID measurements, does not diverge in the presence nor in the absence of a transverse field, showing that this system is neither a ferromagnet nor a spin-glass, in the classical or quantum regimes. Although the possibility of a zero Kelvin transition cannot be excluded, a simple, non-critical, model is proposed to depict the magnetic behavior of the low-doped Ho:LiYF$_4$ crystals.

High-quality LiHo$_x$Y$_{1-x}$F$_4$ crystals with $x=0.165$ and 0.045 were grown in platinum crucibles by the Czochralski method\cite {czo}. The Holmium concentration was accurately determined by x-ray spectral analysis with a CAMEBAX electron-probe micro-analyzer. The magnetization measurements were performed on the micro-SQUID magnetometer\cite{msquid}, between 0.035 K and 0.5K in longitudinal or transverse fields up to 0.5 T. The size of the crystals (330 $\times$ 80 $\times$ 500 $\mu$m$^3$ for the 16.5\% doped) and the field sweep rates (1--50~Oe/s) were small enough to maximize thermal contact and equilibrium with the cryostat. The longest axis of the crystals corresponds to the $c$-axis, which is the easy axis of the magnetization $M$. Additional measurements were performed on a conventional SQUID magnetometer for reference.

\begin{figure}
\includegraphics[width=0.46\textwidth]{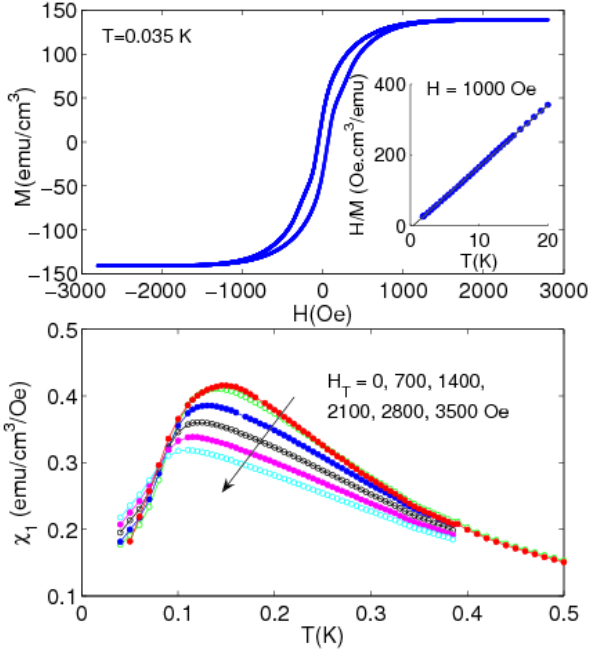}
\caption{(Color online) Top panel: Hysteresis loop of LiHo$_{0.165}$Y$_{0.835}$F$_4$ measured along the easy c-axis at the temperature $T$ = 35 mK at the sweep rate of 35~Oe/s. The insert shows the $T$-dependence of the inverse susceptibility $H/M$ measured in $H$=1000~Oe on a conventional SQUID magnetometer. Bottom panel: Temperature dependence of the initial susceptibility $\chi_1$ of $M(H$) curves for different superimposed transverse magnetic fields $H_T$}
\label{fig-x1}
\end{figure}

The $M(H)$ curves of LiHo$_{0.165}$Y$_{0.835}$F$_4$, measured with the field $H$ applied along the easy $c$-axis at temperatures between 35 and 600 mK, saturate at 6.52 $\mu_b$/Ho. They are hysteretic at low temperatures with a weak S-like shape (see top panel of Fig.~\ref{fig-x1}) suggesting residual phonon bottleneck\cite{chiorescu}, while above 0.18~K they are fully reversible. As seen in the inset, the inverse susceptibility displays a linear $T$-dependence with a paramagnetic Curie temperature $\theta$ = 0.42 K, indicating that ferromagnetic interactions are dominant. Note that a rough evaluation of dipolar interactions yields an energy scale $\sim0.5$~K.

The linear susceptibility $\chi_1$ \cite{phrase} exhibits a broad maximum near 0.16 K, at which $\chi_1$ amounts to $\sim$ 0.41 emu/cm$^3$/Oe (see Fig.~\ref{fig-x1}). This value is not very different from the inverse demagnetization factor of the crystal (shaped as a rectangular prism with the longest dimension parallel to the c-axis) 1/$N$ $\sim$ 0.66. Yet, it is neither much smaller nor comparable to $1/N$, as this is respectively expected for canonical spin-glasses and ferromagnets. There might thus not exist any  long-ranged magnetic order in LiHo$_{0.165}$Y$_{0.835}$F$_4$. In the presence of a transverse field, $\chi_1$ is slightly suppressed, unless the field becomes large. This conventional behavior confirms the absence of ferromagnetic phase transition in this system in zero and finite $H_T$. 

\begin{figure}
\includegraphics[width=0.46\textwidth]{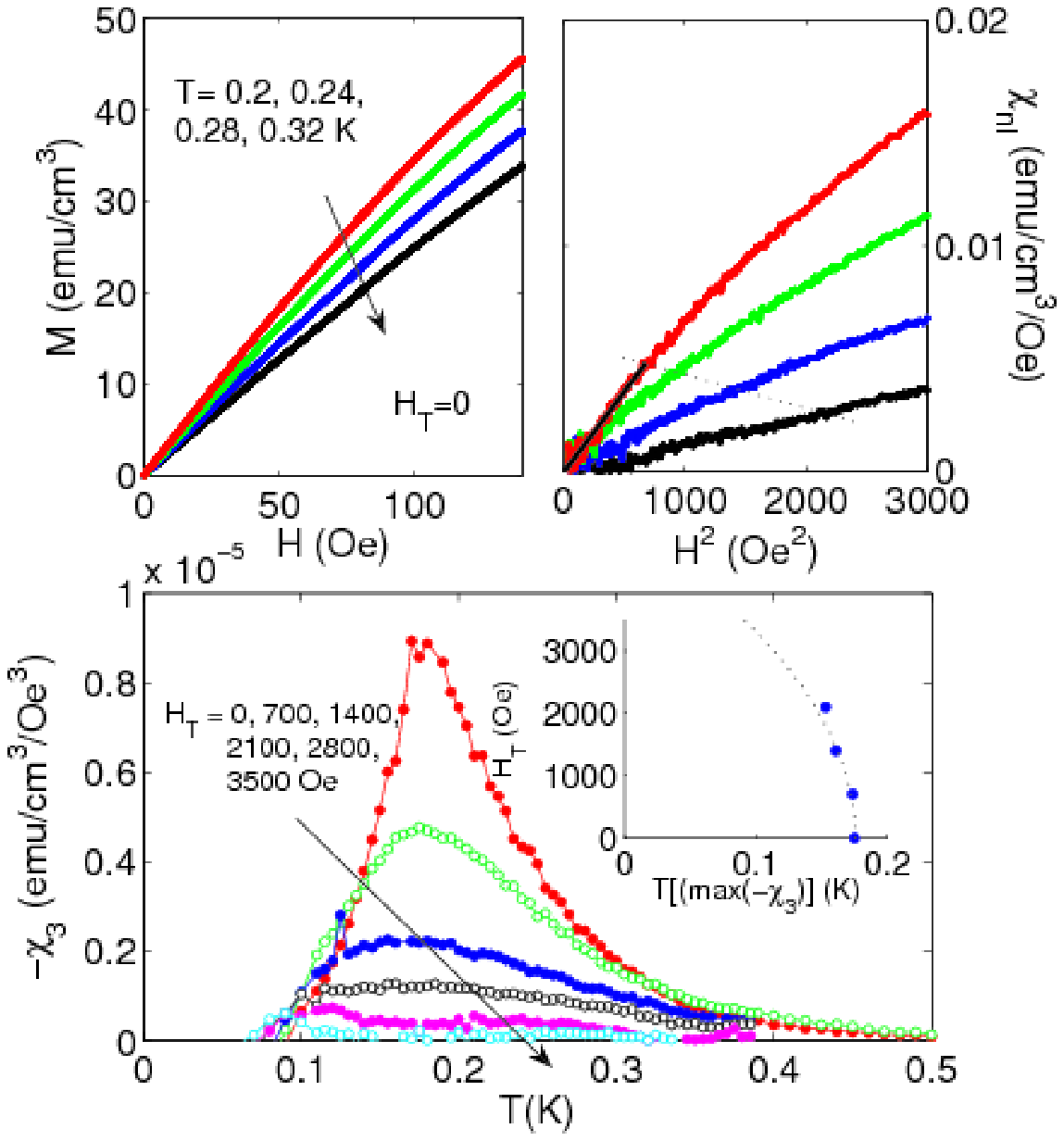}
\caption{(Color online) Top panels: measured $M(H)$ curves or LiHo$_{0.165}$Y$_{0.835}$F$_4$ at selected temperatures (left) and corresponding non-linear susceptibility $\chi_{nl}=\chi_1-M/H$ plotted as a function of $H^2$ for $H_T$=0 and $T$ = 0.2-0.32 K; the initial slopes taken to derive  the $\chi_{3}$ are indicated by thick lines and the dotted line marks the largest $H^2$ values which are  considered (right). Bottom panel: temperature dependence of $\chi_3$ for different transverse fields $H_T$. Note that $\chi_1$  and $\chi_3$  are correctly determined only in the reversible region which is, more or less, above the peak in $\chi_1$. In the low temperature side the ascending and descending branches of the hysteresis loops were averaged in order to approach the metastable equilibrium $M(H)$ curve. The inset displays a $H_T$-$T$ diagram where the $T_{\rm max}$ is the temperature at which the maximum of $-\chi_3$ occurs. Dotted lines are only guides to the eye.}
\label{fig-x3}
\end{figure}

In order to investigate the effect of temperature and transverse field on the non-linear susceptibility, $\chi_{nl}=\chi_1-M/H$ is plotted as a function of $H^2$, as shown in Fig.~\ref{fig-x3} (top panel). This allows us to estimate the lowest order term of $\chi_{nl}$, $\chi_3$, for different transverse fields (see Fig.~\ref{fig-x3}, top panel). Unlike $\chi_1$, $\chi_3$ is greatly affected by small transverse fields ($<$ 70 mT) and is suppressed at larger fields. In zero transverse field, $\chi_3(T)$ exhibits a rather sharp peak, as seem in  Fig.~\ref{fig-x3} (lower panel). Wu et al\cite{aeppli-wu}, who also measured this peak, attributed this sharpness to the divergence of  the spin-glass phase transition. Despite the suppression of $\chi_3$, the same conclusion was reached in the presence of a transverse fields. In their analysis, these authors assumed that $T_g$ corresponds to the susceptibility peak and they erroneously\cite{mattsson} used all the data points available above this temperature, even those where dynamical effects are present, within the peak rounding. In order to clarify this controversial situation, the critical $T$-dependence of $\chi_3$ should be analyzed without making any assumption on the value of $T_g$ and at temperatures slightly above the peak, where $\chi_3$ is fully reversible.

\begin{figure}
\includegraphics[width=0.46\textwidth]{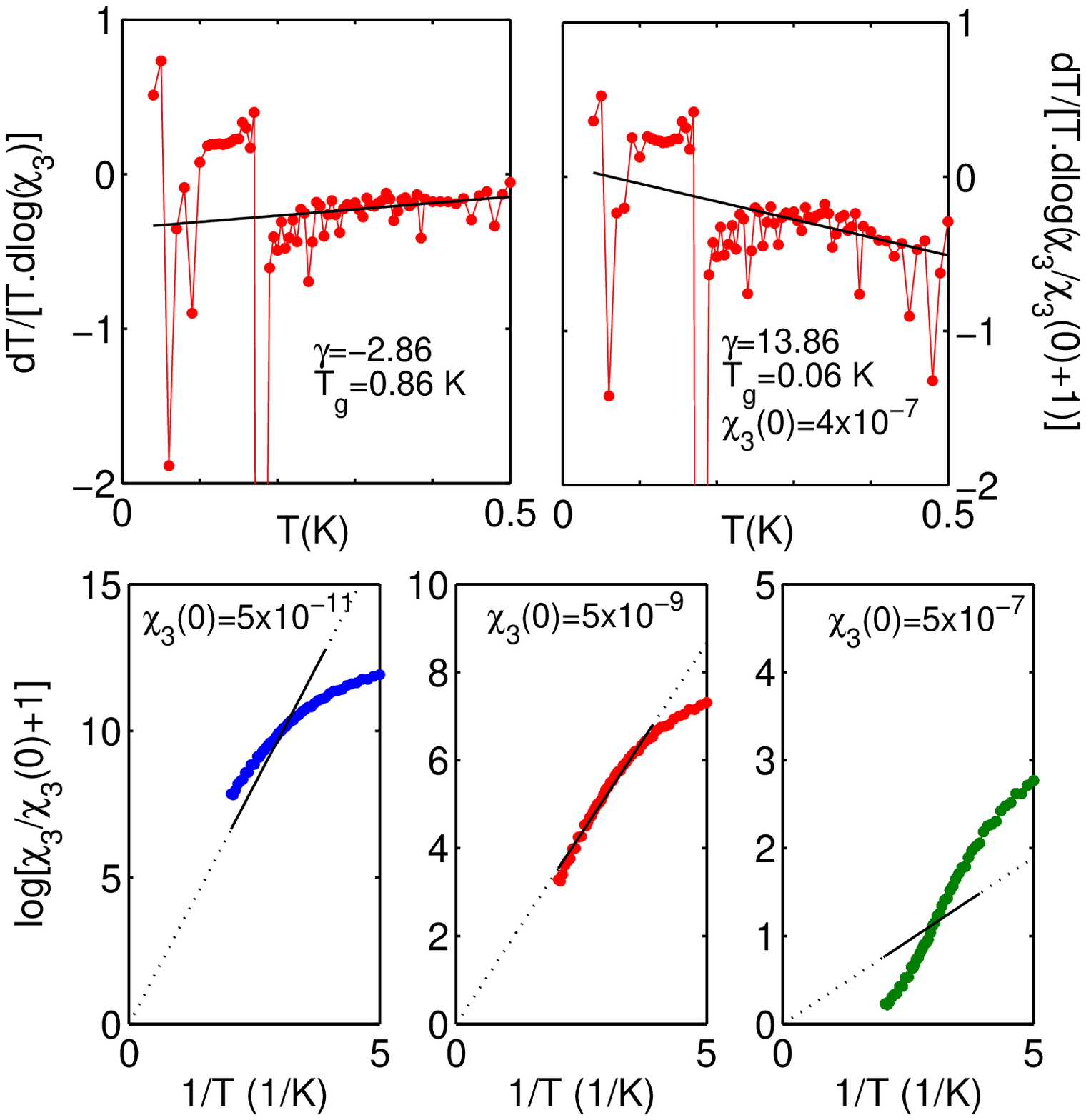}
\caption{(Color online) Analyzes of the $\chi_3$(T) data of LiHo$_{0.165}$Y$_{0.835}$F$_4$ obtained for $H_T$=0. Top panels: plots of  $dT/[T.d\log(\chi_3)]$ (left) and $dT/[T.d\log(\chi_3/\chi_3(0)+1)]$  (right) vs $T$. The straight lines represent linear fits for 0.25 K $<$ $T$ $<$ 0.5 K. Lower panels: plot of $\log(\chi_3/\chi_3(0)+1)$ vs 1/$T$ for different $\chi_3(0)$ values. The straight lines represent linear fits of the data intersecting (0,0) for 0.25 K $<$ $T$ $<$ 0.5 K.}
\label{fig-fits}
\end{figure}

\begin{figure}
\includegraphics[width=0.46\textwidth]{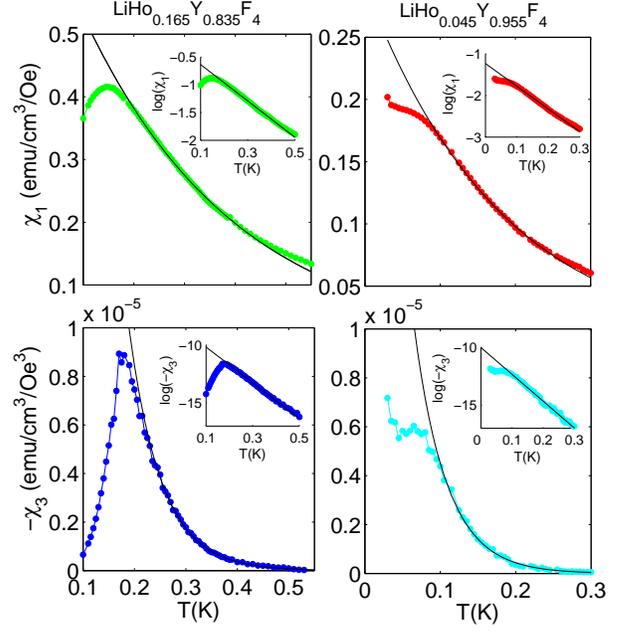}
\caption{(Color online) Main frames: Temperature dependence of the linear and non-linear susceptibilities $\chi_1$ and $\chi_3$ for (left) LiHo$_{0.165}$Y$_{0.835}$F$_4$ and (right) LiHo$_{0.045}$Y$_{0.955}$F$_4$ without transverse field. The continuous lines represent $\exp(-T/T_0)$ fits (see main text), obtained from the $T$-linear fits of $\log(\chi_1)$) and $\log(\chi_3)$) shown in the insets of each panel.}
\label{fig-ccl}
\end{figure}

Phase transition theories and experimental studies of the spin-glass phase transition show that $\chi_3$ should diverge as $\chi_3 \propto [(T-T_g)/T]^{-\gamma}$, where $\gamma$ is a critical exponent\cite{suzuki,barbara}. This implies that $dT/[T.d\log(\chi_3)] =-T/(\gamma T_g)+1/\gamma$, i.e. that $dT/[T.d\log(\chi_3)]$ should be linear with $T$ allowing direct and independent determinations of $\gamma$ and $T_g$ from $T$-linear fits. The plot of the data according to this expression is roughly linear above $\sim$ 0.25K (where $\chi_3$ is fully reversible, see Fig.~\ref{fig-fits}, top left panel). However, the positive slope gives unphysical negative $\gamma$ and much too large $T_g$. Another, slightly improved expression in which $\chi_3$ vanishes at high temperatures was also considered: $\chi_3 = \chi_3(0).([(T-T_g)/T]^{-\gamma} -1)$. The corresponding $dT/[T.d\log(\chi_3/\chi_3(0)+1)]$ was plotted vs $T$ for all possible values of $\chi_3(0)$. When this parameter increases, the slope becomes progressively positive, but the linearity is lost unless $\gamma$ is very large ($\gamma$=13.6 and $T_g$=0.06 K, as exemplified in Fig.~\ref{fig-fits}, top right panel). As the fits based on both expressions, performed without any assumption on $T_g$, lead to unphysical results, we conclude to the absence of a spin-glass phase transition at finite $T_g$ in LiHo$_{0.165}$Y$_{0.835}$F$_4$. The evolution of the maximum of $\chi_3(T)$ with $H_T$ shown in Fig.~\ref{fig-x3} (lower panel), shows clearly the suppression of $\chi_3$. A similar suppression was observed in another series of experiments (not shown) using smaller transverse fields. They both confirm that the maximum value of $\chi_3$ remains finite in zero applied $H_T$. According to the present study, the line $T_g$ vs $H_T$ shown in the inset of Fig.~\ref{fig-x3} (lower panel), is not critical and may simply represent a cross-over between blocked and reversible magnetization when thermal and quantum fluctuations are of the order of long range dipolar interactions (see below). 

Next we investigate the possibility of a phase transition at $T_g$ = 0. A first order expansion of the logarithm of the above form, $\chi_3 = \chi_3(0).([(T-T_g)/T]^{-\gamma} -1)$, was performed for $T_g/T$ $<<$ 1. This yields the characteristic exponential divergence of zero-Kelvin phase transitions: $\log(\chi_3/\chi_3(0)+1)=\gamma T_g/T$. Plots of the data according to $\log(\chi_3/\chi_3(0)+1)$ with $1/T$ are poorly linear for any value of $\chi_3(0)$, except may be for $\chi_3(0)$ $\sim$ 10$^{-9}$ where a rather small linear portion extrapolates to (0,0), as seen in Fig.~\ref{fig-fits}. This suggests that even a zero-Kelvin phase transition is questionable in LiHo$_{0.165}$Y$_{0.835}$F$_4$. However one cannot totally exclude it.

	In order to define the magnetic behavior of the system, one may consider the following, very simple, non-critical approach in which the time-dependent magnetization results from the integration of a uniform distribution of energy barriers with an exponential cut-off at largest energies $E_M$, as $M= \int \exp[-y(E)] dE$, where $y(E)$ = $(t/\tau_0) \exp(-E/k_BT) + E/E_M$. Maximizing the rate of magnetization reversal ($dy(E)/dE$=0) gives the most effective energy scale $E_{\rm eff}$ = $k_BT \log[(E_M/k_BT).(t/\tau_0)]$  ($\sim$ 30 $k_BT$ for quasi-static measurements). This implies that $M_{max}$ $\sim$ $\exp(- y_{max})$ $\sim$ $\exp (-T/T_0)$ where $T_0$ $\sim$ $E_M$/(30$k_B$) represents the measured magnetization. As the magnetic field does not enter explicitly in this expression, it should describe both linear and non-linear susceptibilities (albeit may be with different $T_0$). The $T$-linear fits of $\log(\chi_1)$ and  $\log(\chi_3)$ and resulting $\chi_1(T)$ and $\chi_3(T)$ curves are excellent, as seen in Fig.~\ref{fig-ccl} (left panels). Thus $\chi_1$ and $\chi_3$ appear to follow the above derived $M \sim \exp (-T/T_0)$ form with $T_0$ $\sim$ 0.31 K and 0.063 K respectively.

Similar experiments were also performed on a crystal with lower Ho doping (4.5 \%), i.e. LiHo$_{0.045}$Y$_{0.955}$F$_4$. Except for a temperature shift of the $\chi_3(T)$ curve in the same ratio as the concentrations, the magnetic behavior of this sample was found to be identical to that of the 16.5 \% sample: The maximum value of $\chi_1$ is slightly smaller than 1/$N$; $\chi_3$($T$) does not diverge at finite temperature and is suppressed by a transverse field: scaling plots of the type shown in Fig.~\ref{fig-fits} are also rather poor, indicating the lack of phase transition also in this case. As seen in Fig.~\ref{fig-ccl} (right panels) $\exp(-T/T_0)$ fits are excellent for both $\chi_1(T)$ and $\chi_3(T)$ suggesting as for the 16.5 \% sample, a simple and conventional behavior determined by the highest energy scales. The corresponding values of $T_ 0$  are respectively $\sim$ 0.18 K and 0.043 K, i.e. almost the same as for LiHo$_{0.165}$Y$_{0.835}$F$_4$. The energy scale $E_M$ $\sim$ 30 $k_BT_ 0$, being independent on the concentration, should also be independent of the strength of dipolar interactions. This is not surprising because the relevant energy scale is here determined by hyperfine interactions rather than by dipolar interactions. Indeed, the only way for the magnetization to switch is by thermally activated tunneling on electro-nuclear states distributed over the hyperfine energy, suggesting that $E_M$ $\sim$ 1.8 K. This value is close to the one derived from the concentration independent $T_0$ extracted from the $\chi_3(T)$ curves of both samples, which is $E_M$ $\sim$ 1.6 $\pm$ 0.3 K on average. However the energy derived from the fit of $\chi_1(T)$ is significantly larger (7 $\pm$ 2 K), showing the limits of our simple model.

In conclusion, accurate magnetization measurements performed on single crystals of LiHo$_x$Y$_{1-x}$F$_4$  with $x$=16.5\% and $x$=4.5\% down to 35 mK and their analysis show that both compositions have the same behavior: (i) absence of spin-glass phase transition in a transverse field, as predicted recently\cite{schechter-hyperf,schechter-dip,schechter-new,snider,gingras}, (ii) same absence of phase transition without transverse field as well, at finite temperature; (iii) at zero Kelvin the non-linear susceptibility may be divergent, although the excellent fits of the linear and non-linear susceptibilities of the $\exp(-T/T_0)$ form tend to suggest ordinary thermally activated dynamics in the quantum regime, with $\log(M) = - T f(H,T)$ ($f$ is a functional form). One should note that for low Ho concentrations, Wang-Landau Monte Carlo simulations have predicted a spin-glass transition at zero Kelvin only\cite{snider}. The lack of phase transition at finite temperature (and may be at zero Kelvin as well) should be associated with the disorder inherent to long-ranged dipolar interactions in a diluted system with strong hyperfine interactions.
All these finding are in good agreement with recent theoretical and experimental investigations\cite{schechter-hyperf,schechter-dip,schechter-new,hc,snider,gingras}, and in sharp contrast with earlier studies of LiHo$_x$Y$_{1-x}$F$_4$\cite{aeppli-reich,aeppli-wu,aeppli-science,aeppli-nature}. In particular, the the existence of the so-called antiglass state in the $x=0.045$ sample \cite{aeppli-reich} may now be questioned since both the specific heat \cite{hc} and magnetization (this work) data do not show any marked difference between the $x=0.16$ and $x=0.045$ samples.
 
We thank J. Balay and P. Lejay for their help in shaping the crystals. P. E. J. and R. M. acknowledge the Swedish Research Council (VR) for financial support.  B.B and A.M.T. acknowledge the European contract INTAS-2003/05-51-4943.


\begin{thebibliography}{1}
\bibitem {phrase}
In low fields, the magnetization $M$ can be expanded as $M=\chi_1H+\chi_3H^3+\chi_5H^5+...$ where $\chi_1$ is the linear susceptibility and the non-linear susceptibility $\chi_{nl}=\chi_1-M/H$ contains the higher-order, non-linear, terms.
\end{thebibliography}
\end{document}